\documentclass[review,3p,authoryear]{elsarticle}

\UseRawInputEncoding
\usepackage[multiple]{footmisc}
\usepackage{pdflscape}

\usepackage{amsmath}

\usepackage{enumitem}
\usepackage{rotating} 
\usepackage{lineno}
\usepackage{amssymb}
\usepackage{sidecap}
\usepackage{graphicx}
\usepackage{caption}
\usepackage{subcaption}
\usepackage{multirow}
\usepackage{wrapfig}
\usepackage{xcolor}

\usepackage{lipsum}
\usepackage{nicefrac}
\usepackage{ctable}
\usepackage{booktabs,tabularx}
\usepackage{adjustbox}
\usepackage{floatrow}
\usepackage{color}
\usepackage{xcolor}
\usepackage{colortbl}
\usepackage{float}
\restylefloat{table}
\newcolumntype{x}{>{\raggedright\arraybackslash}X}
\usepackage{bbding}
\usepackage{amsmath}
\usepackage{romannum}
\usepackage[colorlinks=true,
            linkcolor=magenta,
            urlcolor=gray,
            citecolor=cyan]{hyperref}
\newfloatcommand{capbtabbox}{table}[][\FBwidth]
\usepackage{blindtext}
\usepackage{enumitem}
\usepackage{algpseudocode}
\usepackage{xcolor}
\usepackage{adjustbox}
\usepackage{array}
\usepackage[linesnumbered,ruled,vlined]{algorithm2e}
\usepackage[final]{pdfpages}
\usepackage{amssymb}
\usepackage{pifont}
\usepackage{multicol}   
\usepackage{lipsum} 

\usepackage{framed,multirow}
\usepackage{textcomp}
\usepackage{amssymb}
\usepackage{latexsym}

\usepackage{url}
\usepackage{xcolor}

\usepackage[multiple]{footmisc}
\usepackage{pdflscape}
\usepackage{amsmath}

\usepackage{enumitem}
\usepackage{rotating} 
\usepackage{lineno}
\usepackage{amssymb}
\usepackage{sidecap}
\usepackage{graphicx}
\usepackage{caption}
\usepackage{subcaption}
\usepackage{multirow}
\usepackage{wrapfig}
\usepackage{xcolor}
\usepackage{lipsum}
\usepackage{nicefrac}
\usepackage{ctable}
\usepackage{booktabs,tabularx}
\usepackage{adjustbox}
\usepackage{floatrow}
\usepackage{color}
\usepackage{xcolor}
\usepackage{colortbl}
\usepackage{float}
\restylefloat{table}
\newcolumntype{x}{>{\raggedright\arraybackslash}X}
\usepackage{bbding}
\usepackage{amsmath}
\usepackage{romannum}
\usepackage[colorlinks=true,
            linkcolor=magenta,
            urlcolor=gray,
            citecolor=cyan]{hyperref}

\usepackage{blindtext}
\usepackage{enumitem}
\usepackage{algpseudocode}
\usepackage{xcolor}
\usepackage{adjustbox}
\usepackage{array}
\usepackage{graphicx} 
\usepackage{tabularx}
\usepackage{array}
\usepackage{booktabs}
\usepackage{siunitx} 
\usepackage[linesnumbered,ruled,vlined]{algorithm2e}
\usepackage[final]{pdfpages}
\usepackage{amssymb}
\usepackage{pifont}

\usepackage{multicol}   
\usepackage{lipsum}
\usepackage{bbm}
\usepackage{mathtools}
\usepackage{url,longtable,multirow,morefloats,stfloats, floatflt,cancel,tfrupee}

\newcounter{includepdfpage}
\newcounter{currentpagecounter}

\newcolumntype{t}{>{\hsize=0.25\hsize}x}
\newcolumntype{s}{>{\hsize=0.75\hsize}x}
\newcolumntype{X}{>{\hsize=1.5\hsize}x}
\newcolumntype{R}[2]{%
    >{\adjustbox{angle=#1,lap=\width-(#2)}\bgroup}%
    l%
    <{\egroup}%
}

\newcommand{\tabincell}[2]{\begin{tabular}{@{}#1@{}}#2\end{tabular}}

\usepackage{xparse}
\NewDocumentCommand{\rot}{O{90} O{1em} m}{\makebox[#2][l]{\rotatebox{#1}{#3}}}%

\modulolinenumbers[5]
\definecolor{newcolor}{rgb}{.8,.349,.1}
\begin{document}


\begin{frontmatter}

\title{Inter-Rater Uncertainty Quantification in Medical Image Segmentation \\ via Rater-Specific Bayesian Neural Networks}

\author[sustech,ucla]{Qingqiao Hu\textsuperscript{$\ast$\textdagger}}
\author[sustech]{Hao Wang\textsuperscript{$\ast$}}
\author[sustech]{Jing Luo}
\author[sustech,bu]{Yunhao Luo\textsuperscript{\textdagger}}
\author[nanjing]{Zhiheng Zhang}
\author[tum-neur]{Jan S. Kirschke}
\author[tum-neur]{Benedikt~Wiestler}
\author[tum,uzh]{Bjoern~Menze}
\author[sustech]{Jianguo Zhang}
\author[tum,uzh]{Hongwei Bran Li}

\address[sustech]{Research Institute of Trustworthy Autonomous System and Department of Computer Science and Engineering, Southern University of Science and Technology (SUSTech), Shenzhen, China.}
\address[ucla]{Department of Electrical and Computer Engineering, University of California, Los Angeles.}
\address[bu]{Department of Computer Science, Brown University.}
\address[tum]{Department of Computer Science in Technical University of Munich in Germany.}
\address[uzh]{Department of Quantitative Biomedicine in the University of Zurich in Switzerland.}
\address[tum-neur]{Department of Neuroradiology, Klinikum rechts der Isar, Technical University of Munich in Germany.}
\address[nanjing]{Department of Hepatobiliary Surgery, the Affiliated Drum Tower Hospital of Nanjing University Medical School, China.}

\let\thefootnote\relax\footnotetext{\textsuperscript{$\ast$} Qingqiao Hu and Hao Wang contributed equally to this work.}
\let\thefootnote\relax\footnotetext{\textsuperscript{\textdagger} The work was done while Qingqiao Hu and Yunhao Luo was at SUSTech.}

\begin{abstract}
Automated medical image segmentation inherently involves a certain degree of uncertainty. One key factor contributing to this uncertainty is the ambiguity that can arise in determining the boundaries of a target region of interest, primarily due to variations in image appearance. On top of this, even among experts in the field, different opinions can emerge regarding the precise definition of specific anatomical structures.
This work specifically addresses the modeling of segmentation uncertainty, known as inter-rater uncertainty. Its primary objective is to explore and analyze the variability in segmentation outcomes that can occur when multiple experts in medical imaging interpret and annotate the same images. 
We introduce a novel Bayesian neural network-based architecture to estimate inter-rater uncertainty in medical image segmentation. Our approach has three key advancements. Firstly, we introduce a one-encoder-multi-decoder architecture specifically tailored for uncertainty estimation, enabling us to capture the rater-specific representation of each expert involved. Secondly, we propose Bayesian modeling for the new architecture, allowing efficient capture of the inter-rater distribution, particularly in scenarios with limited annotations. 
Lastly, we enhance the rater-specific representation by integrating an attention module into each decoder. This module facilitates focused and refined segmentation results for each rater. 
We also provide an open-source 3D multi-rater dataset with three raters annotated liver lesions in CT images as an additional contribution. This dataset serves as an independent resource for further research in the field.
We conduct extensive evaluations using synthetic and real-world datasets to validate our technical innovations rigorously. Our method surpasses existing baseline methods in five out of seven diverse tasks on the publicly available \emph{QUBIQ} dataset, considering two evaluation metrics encompassing different uncertainty aspects. Furthermore, on the \emph{LIDC-IDRI} dataset, our approach demonstrates superior performance and accurately captures the inter-rater distribution of segmentations.

Our codes, models, and the new dataset are available through our GitHub repository: \hyperref[]{\url{https://github.com/HaoWang420/bOEMD-net}}. 

\end{abstract}


\end{frontmatter}

\section{Introduction}
Accurate medical image segmentation is critical in clinical diagnosis, treatment planning, and various downstream tasks. While deep learning approaches have achieved considerable success in numerous segmentation tasks, most of these tasks are formulated as one-to-one mapping problems \citep{bakas2018identifying,simpson2019large,isensee2021nnu}. Typically, these methods generate a single segmentation result that aligns with a consensus but lacks a measure of segmentation reliability.

\begin{figure}[b]
	\centering
    \includegraphics[width=0.9\textwidth]{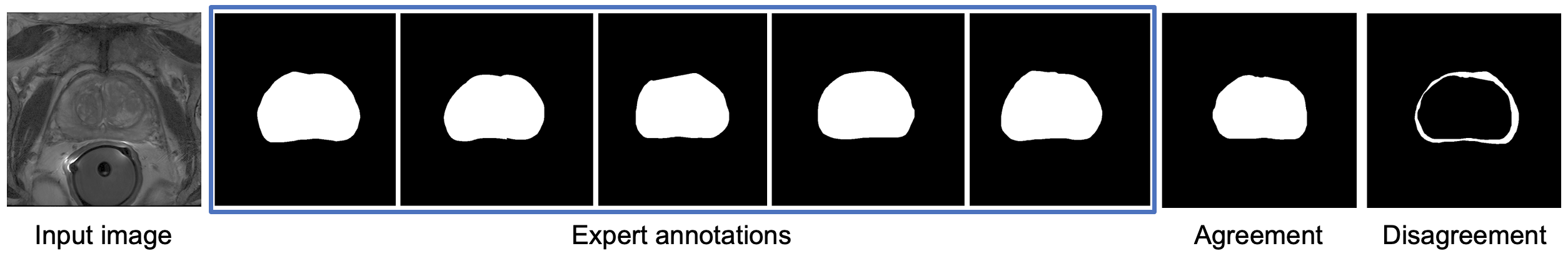}
    \vspace{-0.2cm}
    \caption{A prostate MR image with corresponding annotations from multiple experts showing agreement and disagreement at the pixel level.}
	\label{disagreement}
\end{figure}

However, in practical scenarios, existing image datasets often contain inherent uncertainties due to the ambiguity in the appearances of target structures and variations in expert opinions. Consequently, annotation uncertainties are inevitably introduced, affecting the accuracy and robustness of segmentation algorithms. The presence of inter-rater disagreement is particularly notable, encompassing various structures and pathologies \citep{joskowicz2019inter,becker2019variability}. This can be observed in the example of prostate segmentations depicted in Figure~\ref{disagreement}, where different raters provide divergent annotations, highlighting the challenges posed by such discrepancies.


\begin{figure*}[t]
	\centering
    \includegraphics[width=0.95\textwidth]{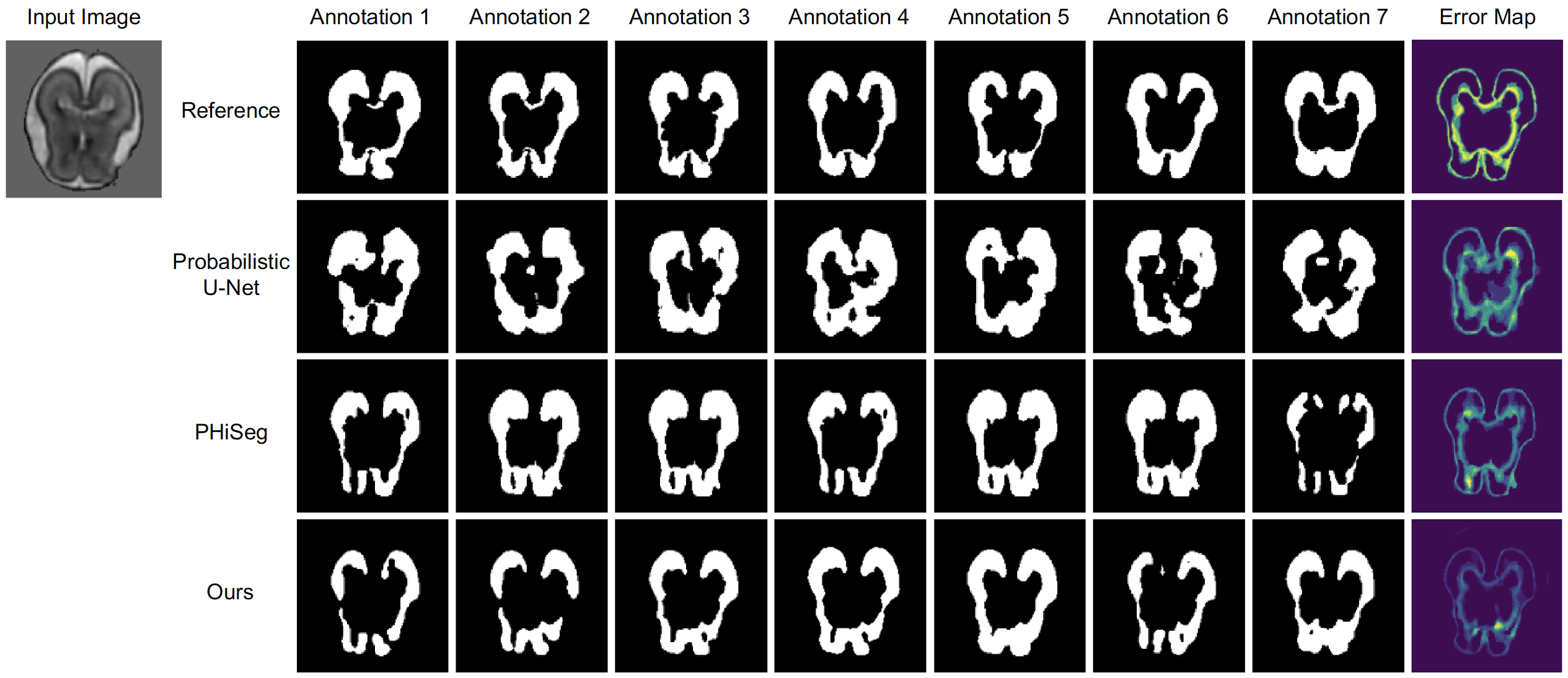}
        \caption{A qualitative comparison was performed among the probabilistic U-Net \citep{kohl2018probabilistic}, {PHiSeg} \citep{baumgartner2019phiseg} and our proposed method. The error map quantifies the difference between predicted and observed distributions. From the segmentation outputs and the error map, one can see that our method captures inter-rater uncertainty more accurately compared to the others.}
	\label{compare}
\end{figure*}

One widely used approach to address inter-rater uncertainty is employing a majority vote method, which considers the most representative opinion among experts as the ground truth. A popular method for modeling the reliability of individual experts and fusing multi-rater labels is \emph{STAPLE} \citep{warfield2004simultaneous}.
However, label fusion-based approaches are limited in their ability to incorporate information from individual annotations. As a result, there is a need for accurate modeling of inter-rater uncertainty to provide a more comprehensive understanding of plausible segmentations given a test image. This entails capturing the range of possible segmentations based on the collective interpretations of the raters, rather than relying solely on a single consensus-based approach. \cite{kohl2018probabilistic} proposed a probabilistic U-Net architecture that could generate unlimited and diverse segmentation masks to reproduce the underlying annotation distribution by sampling the latent space. \cite{baumgartner2019phiseg} improved the output diversity by introducing stochasticity into multiple resolution levels and performing gradual refinement during training.
However, although both methods can generate infinite segmentation with high sample diversity, we observe they do not guarantee accurate modeling of the \emph{inter-rater} uncertainty, as shown in an example in Figure~\ref{compare}. Practically, an ensemble of vanilla U-Nets can better model the inter-rater uncertainty which will be demonstrated in the experiments section. 

We argue that employing a single set of latent variables and the hierarchical architecture proposed in \cite{kohl2018probabilistic} and \cite{baumgartner2019phiseg}, respectively, is insufficient to model an underlying inter-rater uncertainty accurately. We highlight two reasons for this limitation. Firstly, sampling the latent space may lead to poor segmentation outcomes after passing through the decoding layers. This can result in a lack of fidelity in capturing the inherent uncertainty in the inter-rater annotations.
Secondly, it is crucial to consider the annotations from multiple raters as independent sets. Typically, raters perform their annotations separately, and the judgments of one rater do not influence the others. Thus, a more appropriate modeling approach should account for this independence and treat each rater's annotations as distinct sources of information. 
In this work, we focus on learning a generative segmentation model to perform segmentation with high consistency to observed inter-rater distribution without sacrificing the diversity of the predicted segmentations. 


As discussed previously, existing methods like \emph{PHiSeg} estimate different types of uncertainties using a set of correlated latent variables, which may not adequately match the observed distribution due to the independent nature between raters.
In contrast to the design of \emph{PHiSeg}, we propose an extension of the standard U-Net architecture, introducing multiple branches with distinct sets of latent variables, as depicted in Figure~\ref{framework}. Each branch in our architecture aims to capture the uncertainty of an independent subset associated with a specific annotation set.
By incorporating multiple branches, our architecture enables more precise estimation of inter-rater uncertainty. Additionally, we enhance the rater-specific features by introducing the attention mechanisms into the transition connections, facilitating the information transfer from the encoder to the decoders at multiple levels.
Furthermore, we adopt Bayesian modeling, ensuring efficient modeling even with limited training data. Given the absence of uniformly identified raters in existing multi-rater datasets, we have collected a rater-aligned liver tumor dataset, annotated by three experts. This dataset serves to validate the rationale behind our motivation and design choices.
In summary, this work has the following key contributions:

\begin{itemize}[noitemsep]
   \item We propose an extension of the U-Net architecture, transforming it into a \emph{one-encoder-multi-decoder} one. This enhanced architecture enables more precise capture of inter-rater uncertainty in medical image segmentation tasks.
   \item We introduce the concept of \emph{rater-specific attention}, enhancing the skip connections within the network. This attention mechanism facilitates the modeling of rater-specific representations.
   \item We extend the proposed architecture by incorporating Bayesian modeling. This Bayesian modeling approach effectively captures the inter-rater distribution while mitigating the risk of overfitting.
   \item We release two rater-aligned datasets to the medical imaging community to serve as a valuable resource for validating and further advancing research in inter-rater uncertainty modeling.
\end{itemize}


\section{Related Work}

In deep learning models, uncertainty estimation plays a crucial role in quantifying the reliability and robustness of predictions. Uncertainty can generally be categorized into two main types, as outlined by \cite{kendall2017uncertainties}. 
Firstly, \emph{epistemic uncertainty} characterizes the uncertainty inherent in the model itself. This uncertainty arises due to limitations in the training data during the optimization process. 
The second type of uncertainty is \emph{aleatoric uncertainty}, also called data uncertainty. It captures the inherent noise and randomness in the input data. 
Uncertainty estimation has been explored in medical imaging tasks such as super-resolution \citep{tanno2017bayesian} and classification \citep{sudre2019lets} tasks.  


In the context of image segmentation tasks, researchers have explored the use of dropout as a means to estimate \emph{epistemic} uncertainty. 
\cite{nair2018} and \cite{roy2018} employed dropout, as introduced by \citep{gal2016dropout}, to estimate epistemic uncertainty.
The \emph{U2-Net}, proposed by \citep{orlando2019u2}, was specifically designed for photoreceptor layer segmentation in pathological OCT scans. \emph{U2-Net} leverages the {epistemic} uncertainty maps to aid in the manual labeling process.
Considering the scarcity of training data, \cite{seebock2019exploiting} proposed a weakly supervised anomaly detection method. They utilized dropout-based uncertainty estimation and a majority-ray-casting-based post-processing technique to achieve compact segmentation of anomalies. 
\cite{hiasa2019automated} introduced the Bayesian U-Net for muscle segmentation combined with an active-learning framework to reduce manual annotation costs. However, these studies predominantly focused on addressing \emph{epistemic} uncertainty, with limited exploration of aleatoric uncertainty, which captures the inherent variability in the observed data.

\cite{wang2019aleatoric}'s work employed test-time augmentation for estimating \emph{aleatoric} uncertainty. Their approach primarily focused on investigating the impact of different transformations applied to the input images.
In \cite{kendall2017uncertainties}'s study, they induced a Gaussian distribution over the output feature map and estimated each logit's mean and standard deviation. However, since this method did not consider the spatial correlation between pixels, leading to poor quantification of uncertainty \citep{jungo2020analyzing}.
To address this limitation, \cite{monteiro2020stochastic} proposed a novel approach that utilized a Gaussian distribution with a low-rank parameterization of the covariance matrix. By modeling the joint distribution over the output logits, their method generated diverse and plausible outputs that capture the correlation structure. 
In another related work, \cite{zhang2020disentangling} explored methods to estimate the spatial characteristics of labeling errors originating from multiple experts and the distribution of noisy labels. 

From a Bayesian perspective, the probabilistic U-Net \citep{kohl2018probabilistic} is proposed to model complex underlying correlation structures in the segmentation distribution. Combining a conditional variational autoencoder \citep{kingma2014autoencoding} with a U-Net architecture \citep{ronneberger2015u} offers the ability to generate unlimited segmentation outcomes. However, the output diversity of the probabilistic U-Net is limited due to its relatively small sampling space, as pointed out by \cite{baumgartner2019phiseg}.
To overcome this limitation, \emph{PHiSeg} is proposed and uses separated latent variables to govern the variations on multiple resolution levels in a hierarchical and dependent manner. However, it fails to handle high-resolution inputs, as observed by \cite{gantenbein2020revphiseg} in their work. \emph{RevPHiSeg} addresses the memory issue of \emph{PHiSeg} by utilizing reversible blocks. The hierarchical nature of \emph{PHiSeg} and \emph{RevPHiSeg} architectures contradicts the assumption of independence between experts during annotation. 
\cite{bhat2023effect} extends the probabilistic U-Net, allowing more general forms of the Gaussian distribution as the latent space distribution.
Another approach to Bayesian uncertainty estimation in nnU-Net is proposed by \cite{zhao2022efficient}. They introduce an efficient estimation framework based on checkpoints and avoid training multiple models.

Additionally, our experiments reveal that the segmentations generated by \emph{PHiSeg} do not closely align with ground truth annotations. Hence, existing architectures have limitations when it comes to modeling independent multi-rater annotations. These limitations motivate us to address inter-rater uncertainty modeling from both Bayesian and architectural perspectives.



\section{Methodology} 
\subsection{Preliminaries}  \label{pre}
We introduce the definition, the objective of inter-rater uncertainty quantification, and Bayesian modeling and inference. 
\subsubsection{Inter-rater Uncertainty Quantification }
Let $X= \{x_1, x_2, \dots,  x_n\}$ and $Y = \{y_1, y_2, \dots,  y_n\}$ denote a set of $n$ input images and its corresponding set of segmentation maps.  In the multi-rater setting, each image $x_i$ is annotated by $m$ raters, resulting in $m$ segmentation maps, denoted as $y_i = \{y^1, y^2, \dots,  y^m\}$. 
The pixel-level annotation uncertainty (or confidence) can be well represented by a probability ground-truth segmentation map by averaging the $m$ annotations. Without loss of generality to the multi-class scenario, we assume that $y_i$ are binary maps, with "1" representing the foreground and "0" for the background. 
The value of the $j^{th}$ entry in the segmentation map represents its probability belonging to the foreground class, denoted as $p(y_{i, j}=1)$. Hence, the probability segmentation map can be formulated as: 

\begin{equation}
p(y_{i, j}=1) = \frac{1}{m}\sum\limits_{r=1}^{m}y_{i,j}^{r}.
\end{equation}

In other words, it represents the probability that raters agree on the positiveness of pixel $x_{i, j}$. 
Therefore, inter-rater uncertainty quantification aims to predict the probability segmentation map as close as to the ground truth one given input images. 
\subsubsection{Bayesian Inference}
Bayesian neural network (BNN) assumes that the network parameters $w$ follow a prior distribution, i.e., $w \sim P(w)$, which depicts the prior knowledge for a target task. Given the training set $\mathcal{D} = (X, Y)$, 
in the classical setting where only one rater is presented, BNN models the joint distribution of annotations $Y$ and parameters $\textbf{w}$ w.r.t. $X$, denoted as $p(Y, \textbf{w}|X)$ and fits the posterior $p(\textbf{w}|D)$ over network weights $\textbf{w}$: 

\begin{equation}
    \begin{split}
        P(y^*|x^*, \mathcal{D}) &= \int p(y^*|x^*, w)P(\textbf{w}|\mathcal{D})d\textbf{w}, 
    \end{split}
\end{equation}.


In the following sections, we formulate the multi-rater Bayesion modeling and variational inference in a one-encoder-m-decoder architecture.  

\subsection{The One-Encoder-M-Decoder Architecture}
    




Given the training images $X$, let $S_r$ denote the corresponding segmentation maps from $r^{th}$ rater, a subset of $Y$. If we only use $(X, S_r)$ as the training set, this would encapsulate data uncertainty~\citep{kendall2017uncertainties}.When considering all subsets $\{S_r | r = 1, ..., m\}$, modeling the inter-rater distribution would be possible. 
Probabilistic U-Net \citep{kohl2018probabilistic} and \emph{PHiSeg} \citep{baumgartner2019phiseg} both utilize a single-branch architecture to capture inter-rater uncertainty. 
We argue that estimating multiple sources of uncertainties might be overly demanding for a single set of latent variables. 
Intuitively, it might be beneficial to separate the modeling process to correctly capture the distribution of $m$ raters that is independent. Hence, we propose to extend the U-Net architecture from a one-encoder-one-decoder to a one-encoder-m-decoder one, where the $r^{th}$ decoder is responsible for approximating the distribution of $r^{th}$ subset. For a given image, the probability of the distribution is denoted as $p(y_{i}^{r}|x_{i}, S_{r})$, ideally from $r^{th}$ rater. 
The resulting prediction at the $r^{th}$ decoder is defined as $p(\hat{y}_{i}^{r}|x_{i}, B_r)$, where $\hat{y}$ is the predicted label and $B_r$ is the indicator for each branch ($B_r = 1,....,m$). The resulting prediction at each decoder is expected to closely match the underlying distribution $p(y|x, S_r)$. $B_r$ and $S_r$ are not strictly correlated, if rater ids are not aligned. The final prediction is the unified distribution of all branches, which could be formulated as follows:
\begin{equation}
    p(\hat{y}_{i}|x_{i})
    =\sum\limits_{r=0}^{m-1} p(\hat{y}_{i}^{r}|x_{i}, B_r)p(B_r)
\end{equation}
where $p(B_r)$ is the prior distribution of $r^{th}$ decoder. 

\begin{figure}[t]
	\centering
    \includegraphics[width=0.75\textwidth]{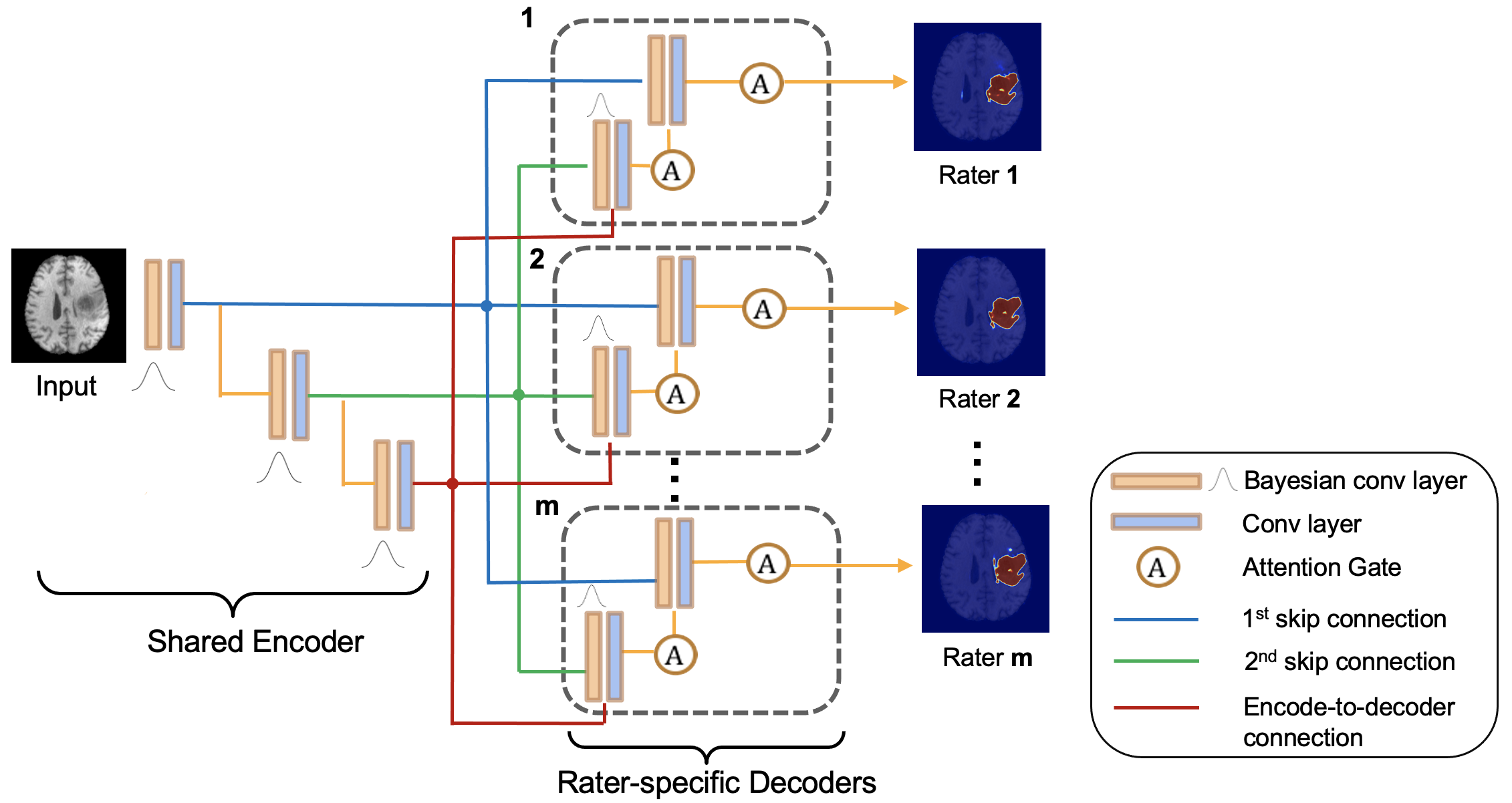}
    
    \caption{A schematic view of our architecture to estimate inter-rater uncertainty with Bayesian modeling. It contains one Bayesian encoder and multiple Bayesian decoders. In the skip connection (red, blue, and green) between the encoder and each decoder, an attention module (denoted by A) is introduced to capture rater-specific representation for individual raters.}
	\label{framework}
\end{figure}

Notably, the encoder is shared for all decoders to extract image features. 
Our \emph{one-encoder-m-decoder} (OM) architecture with Bayesian modeling is shown in Figure~\ref{framework}.
The input data is first fed into the encoder for multi-scale feature extraction followed by $m$ decoders for up-sampling and inference. 
This \emph{OM} architecture offers two advantages: (1) different sources of uncertainties could be handled less correlatedly. (2) the shared encoder enhances the feature representation of the input images when given multiple annotations during training.


\subsection{Multi-rater Bayesian Modeling}
We further introduce Bayesian modeling to our \emph{OM} architecture. The motivation for using Bayesian modeling are two aspects: (1) preventing the network from over-fitting on small datasets, and (2) uncertainty quantification in a Bayesian manner.   
%
Following Section \ref{pre}, the prior probability function of the parameters $w_{r}$ in the decoder branch $B_{r}$ follows the prior distribution $P(w_r)$. 
Given the training set $D$, including all annotation subsets, the inference process of our \emph{OM} architecture with Bayesian modeling can be formulated as:
\begin{equation}
    \begin{split}
        P(\hat{Y}|X, D) &= \int P(\hat{Y}|X, S')P(S'|D)d S' \\
                    &= \frac{1}{m}\sum_{r=1}^{m} P(\hat{Y}|X, S_r)
    \end{split}
\end{equation}
where $\hat{Y}$ is the output segmentation.
Since the probability of each annotation being assigned to each subset $S_r$ is equal, the final predicted distribution $P(\hat{Y}|X, D)$ can be represented as the expectation of the predicted distribution $P(\hat{Y}|X, S_r)$ at each decoder branch. Moreover, $P(\hat{Y}|X, S_r)$ can be formulated as the expectation of each decoder branch under the posterior distribution $P(w_r|S_r)$:
\begin{equation}
        P(\hat{Y}|X, S_r) = \int P(\hat{Y}|X, w_r^{'})P(w_r^{'}|S_r)dw_r^{'} \\ 
        = \mathbbm{E}_{P(w_r|S_r)}[P(\hat{Y}|X, S_r)]
    \label{posterior_equation}
\end{equation}

\subsection{Posterior Distribution Approximation}
From Equation~\ref{posterior_equation}, if the posterior distribution $P(w_r|S_r)$ is known, the predicted output of each branch would be easy to obtain. However, $P(w_r|S_r)$ is numerically expensive to compute, since $p(w_r | S_r) = \frac{P(S_r|w_r) P(w_r)}{P(S_r)} = \frac{ P(S_r | w_r)P(w_r) } {\int P(S_r, w_r^{'})P(w_r^{'})dw_r^{'}}$ needs integration over all the possible weight in each decoder branch. Therefore, an approximation is needed for the posterior distribution $P(w_r|S_r)$. 

To achieve that, variational inference method \citep{blundell2015weight} is used in the training stage. For each decoder branch $B_r$, a distribution $Q(w_r | \theta_j, B_r)$ parameterized by $\theta_j$ is adopted to approximate the true posterior $P(w_r | S_r)$ via minimizing the Kullback-Leibler divergence between $Q(w_r | \theta_jB_r)$ and $P(w_r | S_r)$. The optimization objective function can be formulated as:
\begin{equation}
    \begin{split}
        \theta_{j}^\ast &= \arg \min_{\theta_j} \text{KL}(Q(w_r | \theta_j, B_r) \ || \ P(w_r | S_r))
    \end{split}
    \label{theta_ast}
\end{equation}

The loss function for each branch $B_r$ is:
\begin{equation}
\begin{split}
    \mathcal{L}(B_r) &= \text{KL}(Q(w_r| \theta_j, B                           _j) \ || \ P(w_r | S_r))  + \mathbb{E}_{w_r \sim Q(w_r | \theta, S_r)} \left[\mathcal{L} \left( \hat{y}, \ y \right)\right],
\end{split}
\end{equation}
where $\mathcal{L}(\cdot, \cdot)$ is the Dice loss function. 

The final training loss function of the network is the average loss from each branch $B_r$:
\begin{equation}
\begin{split}
    \mathcal{L} &= \frac{1}{m} \sum_{j=1}^{m} \mathcal{L}(B_r)
\end{split}
\end{equation}


\subsection{Rater-specific Attention}
 The contours of sub-structures, such as the tumor core and edema, may vary among different raters due to individual opinions. To address this, a shared encoder is used to extract shared features at various scales, which are then passed to rater-specific decoders for segmentation.. Traditionally, a skip connection is added in U-Net \citep{ronneberger2015u} to capture the spatial information in early layers. However, naively adding those skip-connections does not allow each rater-specific decoder to have different focuses on the target region. Therefore, we introduce an attention module into each decoder, in a similar fashion to \citep{oktay2018attention}.  

\begin{figure}[t]
	\centering
    \includegraphics[width=0.70\textwidth]{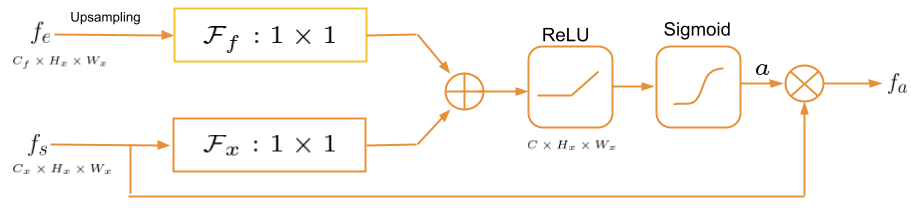}
    \caption{The rater-specific attention module to learn rater-specific features. $f_e$ denotes the feature from the previous layer; $f_s$ denotes the feature from the skip connections. $\mathcal{F}_x$ and $\mathcal{F}_f$ both denote the 1 x 1 convolutional operation. $a$ denotes the attention map, while $f_a$ is the output of attention gate module.}
	\label{attentionGate}
\end{figure}

As shown in Figure~\ref{attentionGate}, the input of the module, $f_e$, is the feature vector captured by the shared encoder, and $f_s$ is the feature vector from the skip-connected layer. Feature vectors $f_e$ and $f_s$ are sent to two 1$\times$1 convolutional layers, respectively to keep them with the same number of channels. Then the outputs from these two 1$\times$1 convolutional layers are added together, going through a sigmoid layer (soft-gating layer). Finally, the input feature $f_e$ will be multiplied by the output $a$ from the sigmoid layer, which is the output $f_a$ of this attention gate unit. The attention module can be denoted as three following equations:
\begin{equation}
    f_g =  \text{ReLU} \left( \mathcal{F}_x(f_s) + \mathcal{F}_f(f_e) \right),
\end{equation}
\begin{equation}
    f_a = f_e \odot sigma(f_g)
\end{equation}
where $\mathcal{F}_x$ and $\mathcal{F}_f$ denote the 1$\times$1 convolutional operation; $\sigma$ is the sigmoid function; $a$ denotes the attention coefficient and $x_a$ is the output of the attention module.

\section{Experiments}

\subsection{Datasets and evaluation metrics} 

\subsubsection{Public multi-rater datasets: QUBIQ and LIDC-IDRI}
These datasets are the publicly released training and validation sets from the QUBIQ challenge. The four datasets include brain tumor, brain growth, kidney, and prostate, with seven diverse segmentation tasks. For each task, a scan corresponds to multiple ground truth annotations experienced clinicians give. The types of the scans and the corresponding number of annotations are shown in Table~\ref{tab:qubiq_detail}. The second category contains a relatively large dataset, LIDC-IDRI \citep{LIDC-IDRI}. It comprises 1018 thoracic CT scans with lesions annotated by four experts.

\subsubsection{A new rater-aligned liver tumor dataset annotated by three experts} 



Since existing public multi-rater datasets involve a mixture of raters but without rater identities, to establish a rater-aligned annotated dataset to verify our assumption, we recruit three radiologists (Z. Z., J. S. K., and B. W.) to manually annotate the liver tumor in a subset of the public CT liver dataset \emph{LiTS} \citep{bilic2023liver}. Thus each case is annotated by the same group of experts. Therefore, it is possible to evaluate the effect of annotations being both aligned and unaligned (as in existing multi-rater datasets) by shuffling the order of the annotations for each case. Figure~\ref{liver_tumor} (a) is one of the examples. 

\subsubsection{A new synthetic dataset by morphological operations} 
\begin{figure}[b]
	\centering
    \includegraphics[width=0.95\textwidth]{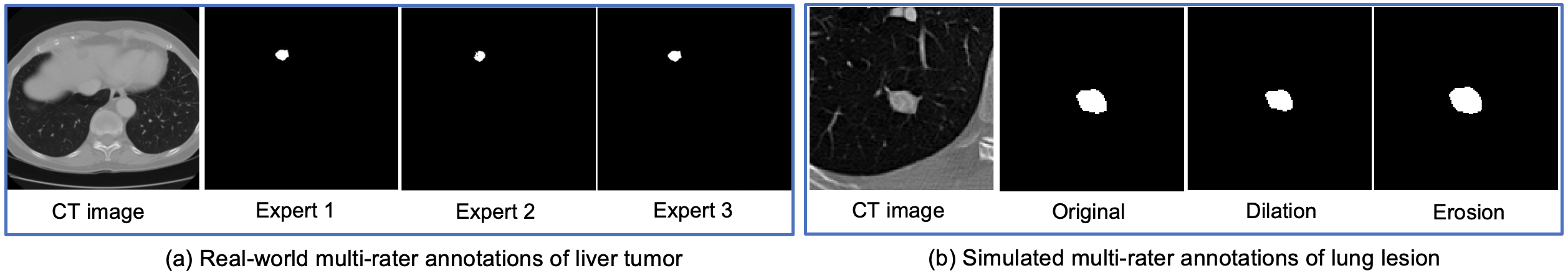}
    \vspace{-0.2cm}
    \caption{To validate our assumption, we contribute (a) a rater-aligned liver tumor dataset annotated by multiple experts and (b) a simulated multi-rater lung lesion segmentation dataset achieved by three morphological operations. }
	\label{liver_tumor}
\end{figure}

Besides the liver tumor dataset, we simulate a synthetic dataset from the public LIDC-IDRI dataset \citep{LIDC-IDRI}. We apply two morphological operations (e.g., erosion and dilation) to the segmentation masks to simulate the data uncertainty.
In our experiment, erosion, dilation, and identity operations are used to obtain three sets of annotations, assuming they are from three raters. One example is shown in Figure~\ref{liver_tumor} (b).

\subsubsection{Implementation}
We implement OMBA-UNet using Pytorch 1.10.1 version \citep{pytorch2019}. We follow the Bayesian-by-Backprop method in \citep{bbb2015} to implement Bayesian layers. These layers are only used for decoders in order to reduce the parameters, since each Bayesian layer has twice the number of parameters than non-Bayesian layers. During the training stage, we use the loss function mentioned above to update the parameters in the network. During the evaluation stage, we output the average of the sampling results. For more implementation details, please refer to our Github repository \url{https://github.com/HaoWang420/bOEMD-net}. 

\subsubsection{Evaluation metrics}
In order to measure the performance of uncertainty estimation of multiple raters for segmentation, 
we primarily adopt the metric\footnote{Note that a simple mean square error (MSE) does not suit here, as 1) it is not designed to measure overlap between segmented regions and target, and 2) it could be easily biased by the large background thus suffering the data imbalance problem compared to the Dice score.} used by the QUBIQ challenge: \url{https://qubiq21.grand-challenge.org/} , termed as Q-score. The Q-score, a staged Dice score, is used to quantify the quality of the predicted probability map $p$ against the ground truth $y$ in $L$ discrete probability levels, formulated as:
\begin{table}[]
  \caption[table: Data Properties]{Summary of the real-world multi-rater datasets used.}
  \label{tab:datasets}
  \centering
  \resizebox{\linewidth}{!}{
  \begin{tabular}{ l | c |c | c | c | c | c }
     \toprule
     \multirow{2}{*}{Task}&  \multicolumn{4}{c|}{QUBIQ} &\multirow{2}{*}{Liver Tumor}  & \multirow{2}{*}{LIDC-IDRI} \\
    \cline{2-5}
       & Brain Tumor & Brain Growth & Kidney & Prostate & &  \\
    \bottomrule
      {\tabincell{l}{Samples(training/test)}}  & 28/4 & 30/5  & 18/4 &34/5 & 1386 (8 patients) / 525 (3 patients) & \\
     \hline
      Number of tasks   & 3 & 1 & 1 & 2 & 1 &1\\
     \hline
     Number of raters &  3 & 7 & 3 & 6 & 3 (rater-aligned) 4\\
     \hline
     Modality &  MRI & MRI & CT & MRI & CT & CT\\
     \hline
     Image size  & $240 \times 240$ & $256 \times 256$ & $497 \times 497$ & $960 \times 960$ &  $400 \times 400$ & $128 \times 128$\\
     \hline
     Number of Input Modalities  & 4 & 1 & 1 & 1 & 1 & 1\\
     \bottomrule
  \end{tabular}}
  \label{tab:qubiq_detail}
\end{table}
\vspace{-0.2cm}
\begin{equation}
    \text{Q-score} = \frac{1}{L}\sum_{l=0}^{L-1}\mathcal{J}\left(T_L(p, l), T_L(y, l)\right),
\end{equation}

For a given discrete probability level $l$, 
\begin{equation}
    T_L(q, l) = 
    \begin{dcases}
        \mathbbm{1}\left\{\frac{l}{L} \leq q < \frac{l + 1}{L}\right\}, & \text{if } 0 \leq l < L - 1 \\
        \mathbbm{1}\left\{\frac{L-1}{L} \leq q \leq 1 \right\}, & \text{if } l = L - 1 \\
    \end{dcases}
\end{equation}
where $\mathcal{J}$ is the Dice score, $\mathbbm{1}$ is a unit step function; $q$ is the continuous probability value; for level $l = L-1$ and $l = l$, they have the same output value.

Compared to the original Dice score, Q-score quantifies the uncertainty by comparing the prediction and ground truth maps at different confidence levels. Since in most of the cases, experts agree on the most parts on the annotations, the variance of different Q-score demonstrates the different ability to capture the annotation uncertainty. 
In addition to Q-score, we use generalized energy distance (GED) which is introduced in \citep{kohl2018probabilistic}, formulated as:
\begin{equation}
    D^2_{GED}(p_{gt},p_s)=2\mathbb{E}[d(s,y)]-\mathbb{E}[d(s,s')]-\mathbb{E}[d(y,y')],
\end{equation}
where $s,s'$ are samples from predicted distribution $p_s$, $\bar{s}$ is the mean prediction, $y,y'$ are samples from ground truth distribution and $d(\cdot, \cdot)=1-\text{IoU}(\cdot, \cdot)$. $D^2_{\text{GED}}$ measures variance of predicted distribution and consistency between predicted distribution and ground truth distribution. 

To specifically quantify the diversity of the outcome by different methods, we adopt \emph{Sample Diversity} used in \citep{baumgartner2019phiseg}, formulated as $\mathbb{E}[d(s, s')]$.  
It describes the average distance between each pair of predicted samples.
Moreover, to measure the similarity between predicted samples and ground truth samples, we define \emph{Sample Similarity} as: $1 - \mathbb{E}[d(s,y)]$.

\subsection{Experiment setup}
\subsubsection{Baselines and parameters setting}

We compare our method with several baselines as follows: \\
1) \textit{Probabilitic U-Net} \citep{kohl2018probabilistic} \\
2) \textit{PHiSeg} \citep{baumgartner2019phiseg} \\
3) \textit{Vanilla U-Net}: The one-encoder-one-decoder U-Net trained with the probability map. \\
4) \textit{U-Net ensemble\footnote{Please refer to the material presented in the challenge workshop: \url{https://qubiq.grand-challenge.org/Home/}. The winner of QUBIQ challenge used an ensemble of multiple U-Nets.}}: An ensemble of several U-Nets trained with one rater independently; the final output is the average of the outputs achieved by all U-Nets. \\
5) \textit{OM-UNet}: A one-encoder-m-decoder U-Net based architecture without Bayesian modeling and attention gates. \\
6) \textit{OMA-UNet}: An \emph{OM-UNet} with attention gates within the skip connection. \\
7) \textit{OMBA-UNet}:  An \emph{OMA-UNet} with Bayesian modeling, which is our proposed method. \\

\subsubsection{Image pre-processing and training setting}
For brain-tumor, brain-growth and kidney tasks in QUBIQ dataset, images are normalize to be in the range of $(-1, 1)$. For the prostate task, images are cropped to be size $640 \times 640$ to concentrate on the region of interest. For LIDC-LDRI dataset \citep{LIDC-IDRI}, we use its pre-processed version described in \citep{kohl2019hierarchical}.




For the \emph{four} datasets from QUBIQ challenge, except \emph{PHiSeg} and probabilistic U-Net, which are optimized based on the internal validation set, other networks are trained for each task with a batch size of 4 and 200 epochs. The training process uses a polynomial learning rate scheduler and SGD \citep{Robbins2007ASA} optimizer with an initial learning rate set to 0.001. We use random 20\% data within the training set as an internal validation set to optimize the hyper-parameters using Q-score. We modify the code repository in \citep{baumgartner2019phiseg} for the Tensorflow version of Probabilistic U-Net on QUBIQ dataset. For more details, please refer to (\hyperref[]{\url{https://github.com/WinstonHuTiger/PHiSeg-code}}).

As for the LIDC-IDRI \citep{LIDC-IDRI} dataset, we train \emph{PHiSeg} and probabilistic U-Net using Adam \citep{adam2015} optimizer with initial learning rate $1 \times 10^{-3}$ until the models converge; for OMBA-UNet, we train the model using Polynomial learning rate scheduler and Adam \citep{adam2015} with initial learning rate $5 \times 10^{-3}$ for 200 epochs; we train other baselines using Polynomial learning rate scheduler and SGD \citep{Robbins2007ASA} with initial learning rate $1 \times 10^{-3}$ for 200 epochs.


\section{Results}

We first verify our basic assumption and the proposed \emph{OM-UNet} architecture on the rater-aligned dataset introduced in previous section. The results are shown in Table~\ref{tab:synthetic}.
Then we present the results achieved by our method on common multi-rater datasets, i.e., QUBIQ and LIDC-IDRI, and compared to state-of-the-art methods in Table~\ref{tab:results_Dice} and Table~\ref{tab:LIDC} respectively. In the following sections, we present and discuss the results from several points of view.




\begin{table}[]
    \centering
    \caption{Results of different architectures and input data on the rater-aligned liver tumor data and the synthetic LIDC-IDRI data under different evaluation metrics. }
  \resizebox{\textwidth}{!}{
    \begin{tabular}{c|c| c|c|c|c |c|c|c|c }
        \toprule
        \multicolumn{2}{c|}{Metrics} & \multicolumn{4}{c|}{rater-aligned liver tumor} & \multicolumn{4}{c}{LIDC-IDRI} \\
        \cline{1-10}
       Multi-decoder  &  Shuffled annotations  & Q-score $\uparrow$ & $D^2_{GED}$ $\downarrow$  &  Diversity $\uparrow$  &  Similarity $\uparrow$ & Q-score $\uparrow$ & $D^2_{GED}$ $\downarrow$  &  Diversity $\uparrow$  &  Similarity $\uparrow$\\
        \midrule
         \checkmark  & \checkmark      & 0.828 & 0.216 & 0.000 & 0.216  & 0.838   & 0.532        & 0.195            & 0.459  \\
        \hline
               \checkmark         &     -   & \textbf{0.859}   & \textbf{0.195}        & \textbf{0.125} & \textbf{0.268}  & \textbf{0.844}   & \textbf{0.519}        & \textbf{0.279}            & \textbf{0.494} \\
        \hline
          -   &      -        & 0.828 & 0.216 & 0.000 & 0.216     & 0.838   & 0.706        & 0.020            & 0.470 \\
        \bottomrule
    \end{tabular}}
    \label{tab:synthetic}
\end{table} 

{
\renewcommand{\arraystretch}{1.2}
\begin{table*}
\centering
\caption{Performances of the methods under Q-score and $D^2_{GED}$ on QUBIQ dataset. Our method achieves the best Q-scores and  $D^2_{GED}$ on five tasks. $*$ indicates reproduced by ourself, and the codes are available.}
\label{tab:results_Dice}
\resizebox{\linewidth}{!}{
\begin{tabular}{c|c|c|c|c|c|c|c|c|l} 
\toprule
                \multirow{2}{*}{} &\multirow{2}{*} {Methods} & \multicolumn{3}{c|}{Brain Tumor}                    & Brain Growth    & Kidney          & \multicolumn{2}{c|}{Prostate}     & \multirow{2}{*}{Average}  \\ 
\cline{3-9}
                  &                          & task1           & task2           & task3           & task1           & task1           & task1           & task2           &                                 \\ 
\hline
\multirow{5}{*}{\rotatebox[origin=c]{90}{Q-score $\uparrow$}} & Prob. U-Net (*)  \citep{kohl2018probabilistic} & 0.3921& 0.5272 & 0.7905 & 0.4428 & 0.7894 & 0.4886& 0.5417 & 0.5675  \\ 
\cline{2 -10} & PHiSeg (*)  \citep{baumgartner2019phiseg} & 0.8182 & 0.7988 & 0.7597 & 0.4502 & \textbf{0.8500} & \textbf{0.5659}  & 0.5019 & 0.6778  \\

\cline{2-10 } & Vanilla U-Net            & 0.8091          & 0.8841          & 0.7887          & 0.4152          & 0.8196          & 0.5227          & 0.5099          &            0.5277           \\ 
\cline{2-10}
                  & U-Net ensemble           & 0.8044          & 0.8970          & 0.7750          & 0.4330          & 0.8207          & 0.5292          & 0.5236          &     0.5314                            \\ 
\cline{2-10}
                  & {OM-UNet}              & 0.8131          & 0.8905          & 0.7947 & 0.4472          & 0.8221          & 0.5431 & 0.5256          &    0.6793                             \\ 
\cline{2-10}
                  & OMA-UNet            & 0.8137 & 0.9035          & 0.7442          & 0.4542          &0.8315          & 0.5353          & 0.5255          &        0.6868                         \\ 
\cline{2-10}
                  & OMBA-UNet (ours)          & \textbf{0.8222}         & \textbf{0.9093} & \textbf{0.8333}          & \textbf{0.4732} & 0.8278 & 0.5362          & \textbf{0.5852} &     \textbf{0.7125}                            \\ 
\midrule
\multirow{5}{*}{\rotatebox[origin=c]{90}{$D^2_{GED}\downarrow$}}  & Prob. U-Net & 0.1848 & 0.4134 & 0.4896& 0.2038 & 0.2714 & 0.2163 & 0.3129 & 0.2989 \\
\cline{2 -10} & PHiSeg & 0.2125 & 0.7230 & 0.8321 & 0.2279 & \textbf{0.1002} & \textbf{0.1149}  & 0.4800 & 0.3844  \\
\cline{2-10} & Vanilla U-Net            & 0.2428          & 0.3465          & 0.7251          & 0.2442          & 0.3916          & 0.2791          & 0.3763          &    0.3722                             \\ 
\cline{2-10}
                  & U-Net ensemble           & 0.2423          & 0.3315          & 0.5386          & 0.2483          & 0.3273          & 0.2350          & 0.2797 &      0.3147                          \\ 
\cline{2-10}
                  & {OM-UNet}               & 0.2257          & 0.3576          & 0.5638          & 0.2138          & 0.3640          & 0.2058 & 0.2840          &     0.3164                            \\ 
\cline{2-10}
                  & OMA-UNet            & 0.2093 & 0.2214 & 0.5022          & 0.1849          & 0.2180 & 0.2321          & 0.3252          &     0.2704                         \\ 
\cline{2-10}
                  & OMBA-UNet (ours)           & \textbf{0.1469}         & \textbf{0.1503}          & \textbf{0.2577} & \textbf{0.1469} & 0.3029          & 0.3139          & \textbf{0.2703}          &   \textbf{0.2616}                              \\
        
\bottomrule
\end{tabular}}
\end{table*}
}

{
\renewcommand{\arraystretch}{1.2}
\begin{table*}[]
\centering
\caption{The model comparison on LIDC-IDRI dataset under Q-score, $D^2_{GED}$, Sample Diversity and Sample Similarity. $*$ indicates the methods re-implemented by our own.}
\begin{tabular}{c|c|c|c|c}
\toprule
Method         & Q-score $\uparrow$ & $D^2_{GED}$ $\downarrow$ &  Diversity $\uparrow$ &  Similarity $\uparrow$             \\ \midrule
Prob. U-Net (*) \citep{kohl2018probabilistic} & 0.6415 & 0.5211 & \textbf{0.4869} & 0.2994 \\ \hline
PHiSeg (*) \citep{baumgartner2019phiseg}         & 0.7067   & 0.3861  & 0.4021 & 0.4093 \\ \hline
 Vanilla U-Net          & 0.6512   & 0.7037           & 0.04573 & 0.4287  \\ \hline
U-Net Ensemble & 0.6985   & 0.4417           & 0.3499 & 0.4076   \\ \hline
OM-UNet  & 0.6876 & 0.4622 & 0.4244 &  0.3601 \\ \hline
OMA-UNet   & 0.6939   & 0.4169           & 0.3281  & 0.4309  \\ \hline
OMBA-UNet (ours) & \textbf{0.7576}   & \textbf{0.3750}            & 0.2915 & \textbf{0.4701}   \\ \bottomrule
\end{tabular}
\label{tab:LIDC}
\end{table*}
}


\subsection{Effectiveness of OM-UNet architecture} 
We first verify our basic assumption: the \emph{OM-UNet} architecture can model rater-specific distributions as it learns to perform segmentation independently in each decoder. 
We examine the performances on both the liver tumor dataset and the synthetic dataset under two conditions as shown in Table~\ref{tab:synthetic}. 
In both tables, all the metrics including Q-score, $D^2_{GED}$, sample diversity and sample similarity are significantly improved compared to single-decoder version. When present with rater-aligned data, i.e. each subset contains a specific annotation from one distinct rater, multi-decoder method better captures inter-rater uncertainty and demonstrates better performance. 
In both Table~\ref{tab:synthetic}, we also observe that shuffling the annotations for the multiple decoders will degrade the general performance. 
This observation indicates that the proposed architecture is rater-specific.  
In Table~\ref{tab:synthetic}, when training with the shuffled data, our method still outperform the single-decoder version on two metrics, i.e., $D^2_{GED}$ and sample diversity.
It indicates that \emph{OM-UNet} architecture can handle the real-world multi-rater data, of which subset/rater specific decoders capture both the subset information under both non-aligned and aligned information.

On the real-world datasets, we observe that the \emph{OM-UNet} architecture outperforms both vanilla U-Net and U-Net ensemble in most of the tasks as shown in Table~\ref{tab:results_Dice} and Table~\ref{tab:LIDC}.
For vanilla U-Net, due to its deterministic behavior and limited capacity, it lacks the ability to estimate the inter-rater ground truth distribution. 
As for the ensemble of multiple U-Net, although it has multiple independent networks for each subset, it does not share the low-level en-visionary information across networks. 
Thus, the ensemble method may end up with sub-optimal features for tasks with limited training data samples, e.g. the brain tumor datasets in QUBIQ. 
In contrast, our proposed \emph{OM-UNet} architecture shares the feature-extractor and dedicated decoders for each subset, thus manages to model inter-rater distributions. 

\subsection{Effectiveness of rater-specific attention.} 
When comparing \emph{OMA-UNet} to \emph{OM-UNet} in Table~\ref{tab:results_Dice}, the former achieves slightly better average Q-score (0.6868 \emph{vs.} 0.6793) and lower average $D^2_{GED}$  (0.2704 \emph{vs.} 0.3164), which demonstrates that attention gates improves the representation at each branch to estimate the subset-specific distribution. 
In Table~\ref{tab:LIDC}, the observation is quite similar that \emph{OMA-UNet} outperforms \emph{OM-UNet} on both Q-Score and $D^2_{GED}$. Notably probabilistic U-Net achieved the high sample diversity which is its main objective.
More importantly, the model with attention gates achieves much higher sample similarity compared to the one without (0.4309 \emph{vs.} 0.3601) in Table~\ref{tab:LIDC}. 
These observations indicate that the attention gates effectively enhances the intermediate representation. These suggests that the attention gates does help the network capture the inter-rater uncertainty. 
\subsection{Effectiveness of Bayesian modeling.}
To demonstrate the effectiveness of Bayesian modeling, we conduct a comparison between \emph{OMA-UNet} and \emph{OMBA-UNet} on the real-world datasets QUBIQ and LIDC-IDRI. 
In Table~\ref{tab:results_Dice}, we observe that Bayesian modeling further improve the efficacy on both Q-score and $D^2_{GED}$. In Table~\ref{tab:LIDC}, our model outperforms the PHiSeg \citep{baumgartner2019phiseg} and probabilistic U-Net \citep{kohl2018probabilistic}.
Moreover, in Table~\ref{tab:LIDC}, we notice that Bayesian modeling increase the sample similarity by a large margin (from 0.4309 to 0.4701) while the sample diversity decreases (from 0.3281 to 0.2915). Although probabilistic U-Net achieved the high sample diversity because it samples the latent space, we argue that the high diversity does not necessarily reflect the property of underlying distribution. Intuitively, as shown in previous results (e.g. Figure~\ref{compare}), the distribution generated by probabilistic U-Net does not match well with the ground-truth distributions. Quantitatively, in Table~\ref{tab:LIDC}, the sample similarity (0.2994) is much lower than other models. 
From both Table~\ref{tab:results_Dice} and Table~\ref{tab:LIDC}, we can observe that the performance gain from Bayesian modeling is relatively more noticeable under small datasets (except kidney dataset). This indicates that Bayesian modeling reduces the over-fitting problem on small datasets. It further suggests that Bayesian modeling well captures the inter-rater distribution from the whole dataset distribution on these small datasets, although the improvement on relatively larger datasets, like LIDC-IDRI is not that significant. These observations indicate that Bayesian modeling accurately captures the inter-rater distribution. 

\subsection{Similarity \emph{vs.} diversity.}
To demonstrate the trade-off between similarity and diversity in  modeling uncertainty, we compare the performance of three models, i.e., probabilistic U-Net, PHiSeg and \emph{OMBA-UNet} on QUBIQ and LIDC-IDRI. 
In Table~\ref{tab:results_Dice}, probabilistic U-Net produces lower Q-score and $D^2_{GED}$, which could be resulted from the limited similarity of probabilistic U-Net samples to the ground truth samples.
Results on LIDC-IDRI dataset proves our assumption. In Table~\ref{tab:LIDC}, it is shown that although probabilistic U-Net generates segmentation samples with high diversity score, its similarity between the prediction and the ground truth is limited as reflected by the similarity metric. 
PHiSeg generally improves probabilistic U-Net by sampling from a relatively larger and hierarchical latent space. 
However, PHiSeg and probabilistic U-Net both suffer from the limited sample similarity as a result of not fully utilizing the subset-specific information.
In Table~\ref{tab:results_Dice} and Table~\ref{tab:LIDC}, by comparing \emph{OMBA-UNet}to PHiSeg and probabilistic U-Net, we find that our method achieves the best performance under both average $D^2_{GED}$ and average Q-score. 
As shown in Table~\ref{tab:LIDC}, among these three models, our method obtains much higher sample similarity with slightly lower sample diversity. It indicates that our method ensures the similarity while maintaining the diversity learned from inter-rater distribution at a high level.

\begin{table*}[h!]
\centering
\caption{The comparison between the four major components that potentially contribute to the performance. Each component, including Ensemble, Bayesian layers (BL), attention gate (AG), one-encoder-multiple-decoder (OM) is listed and each architecture with any specific component is marked with a check mark.}
\resizebox{\linewidth}{!}{
\begin{tabular}{c|c|c|c|c|c|c|c|c|c|c|c|c}
\toprule
\multirow{2}{*}{}        & \multicolumn{4}{c|}{Methods}                              & \multicolumn{3}{c|}{Brain Tumor} & Brain Growth & Kidney & \multicolumn{2}{c|}{Prostate} & \multirow{2}{*}{Average} \\ \cline{2-12}
                         & Ensemble     & BL           & AG           & OM         & task1     & task2     & task3    & task1        & task1  & task1         & task2         &                          \\ \hline
\multirow{4}{*}{Q-score} & $\checkmark$ &              &              &              & 0.8044    & 0.8970    & 0.7750   & 0.4430       & 0.8207 & 0.5292        & 0.5236        & 0.5314                   \\ \cline{2-13} 
                         & $\checkmark$ & $\checkmark$ &              &              & 0.8124    & 0.8673    & 0.8078   & 0.4639       & 0.8264 & 0.5514        & 0.5716        & 0.7001                   \\ \cline{2-13} 
                         & $\checkmark$ & $\checkmark$ & $\checkmark$ &              & 0.8136    & \textbf{0.9180}    & 0.8072   & 0.4689       & 0.8281 & 0.5460        & 0.5757        & 0.7082                   \\ \cline{2-13} 
                         &              & $\checkmark$ & $\checkmark$ & $\checkmark$ & \textbf{0.8222}    & 0.9093    & \textbf{0.8333}   & \textbf{0.4732}       & 0.8278 & 0.5362        & \textbf{0.5804}        & \textbf{0.7125}                   \\ \hline
\multirow{4}{*}{GED}     & $\checkmark$ &              &              &              & 0.2423    & 0.3315    & 0.5386   & 0.2483       & 0.3273 & 0.2350        & 0.2797        & 0.3147                   \\ \cline{2-13} 
                         & $\checkmark$ & $\checkmark$ &              &              & 0.2088    & 0.4358    & 0.6922   & 0.1563       & 0.3723 & \textbf{0.1816}        & 0.2801        & 0.3324                   \\ \cline{2-13} 
                         & $\checkmark$ & $\checkmark$ & $\checkmark$ &              & \textbf{0.1440}     & \textbf{0.2487}    & 0.6423   & 0.1501       & \textbf{0.2287} & 0.1842        & \textbf{0.2606}        & 0.2662                   \\ \cline{2-13} 
                         &              & $\checkmark$ & $\checkmark$ & $\checkmark$ & 0.1503    & 0.2577    & \textbf{0.3894}   & \textbf{0.1469}       & 0.3029 & 0.3139        & 0.2703        & \textbf{0.2616}                   \\ \bottomrule
\end{tabular}}
\label{tab:ensemble_compar}
\end{table*}

\subsection{Visualization}
\begin{figure}[htpb]
	\centering
    \includegraphics[width=0.5\textwidth]{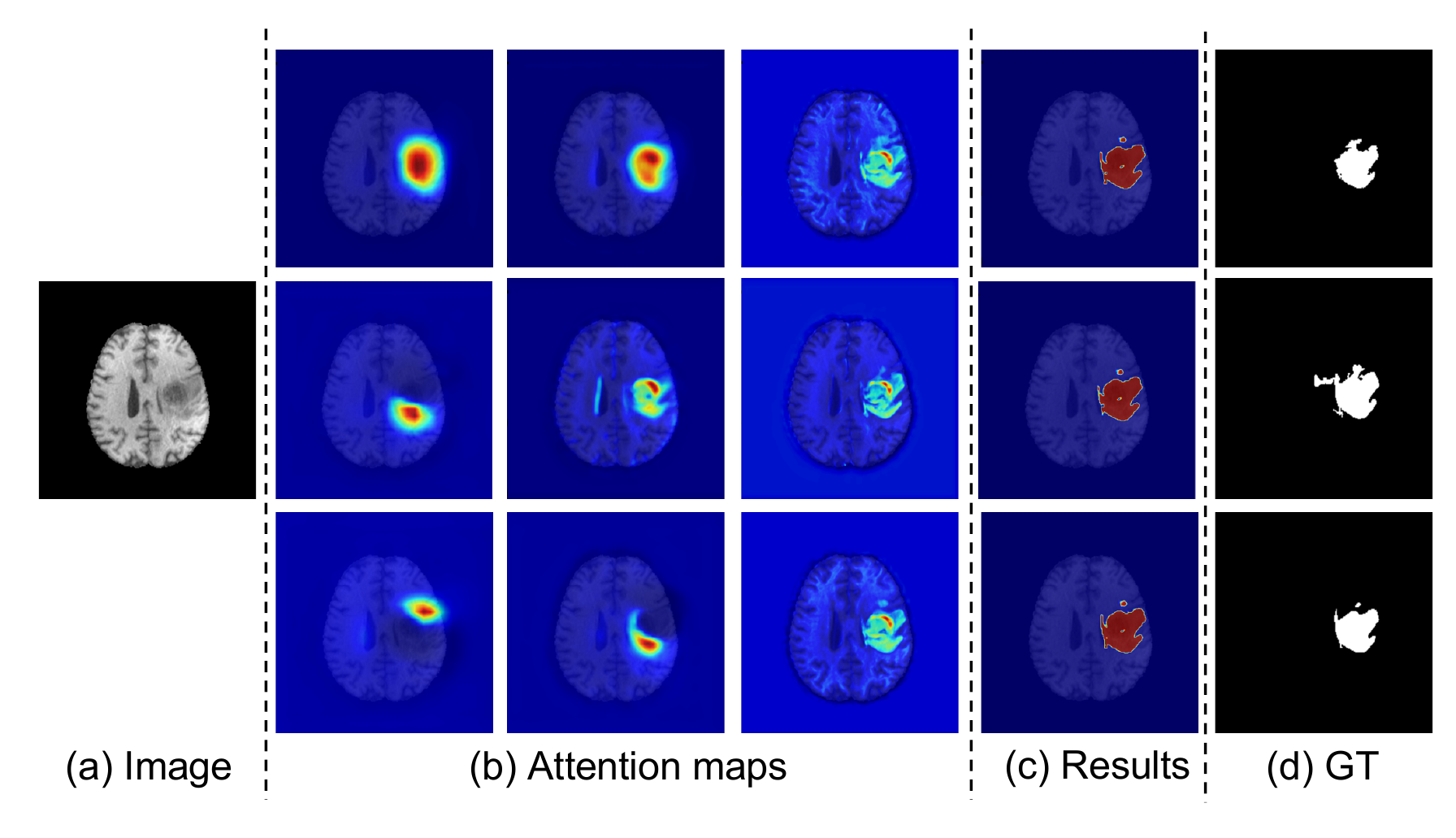}
    \caption{Illustration of the learned attention coefficients of AGs. (a) Sample image showing a brain tumor (Task 1 in QUBIQ). (b) Attention maps: From left to right, the attention maps of AGs in sequenced layers of one decoder. From top to bottom, the attention maps in different decoders. (c) Combined segmentation map by the model. (d) GT: combined annotations by different raters.}
	\label{attention}
\end{figure}

\begin{figure*}
	\centering
    \includegraphics[width=0.95\textwidth, angle=0]{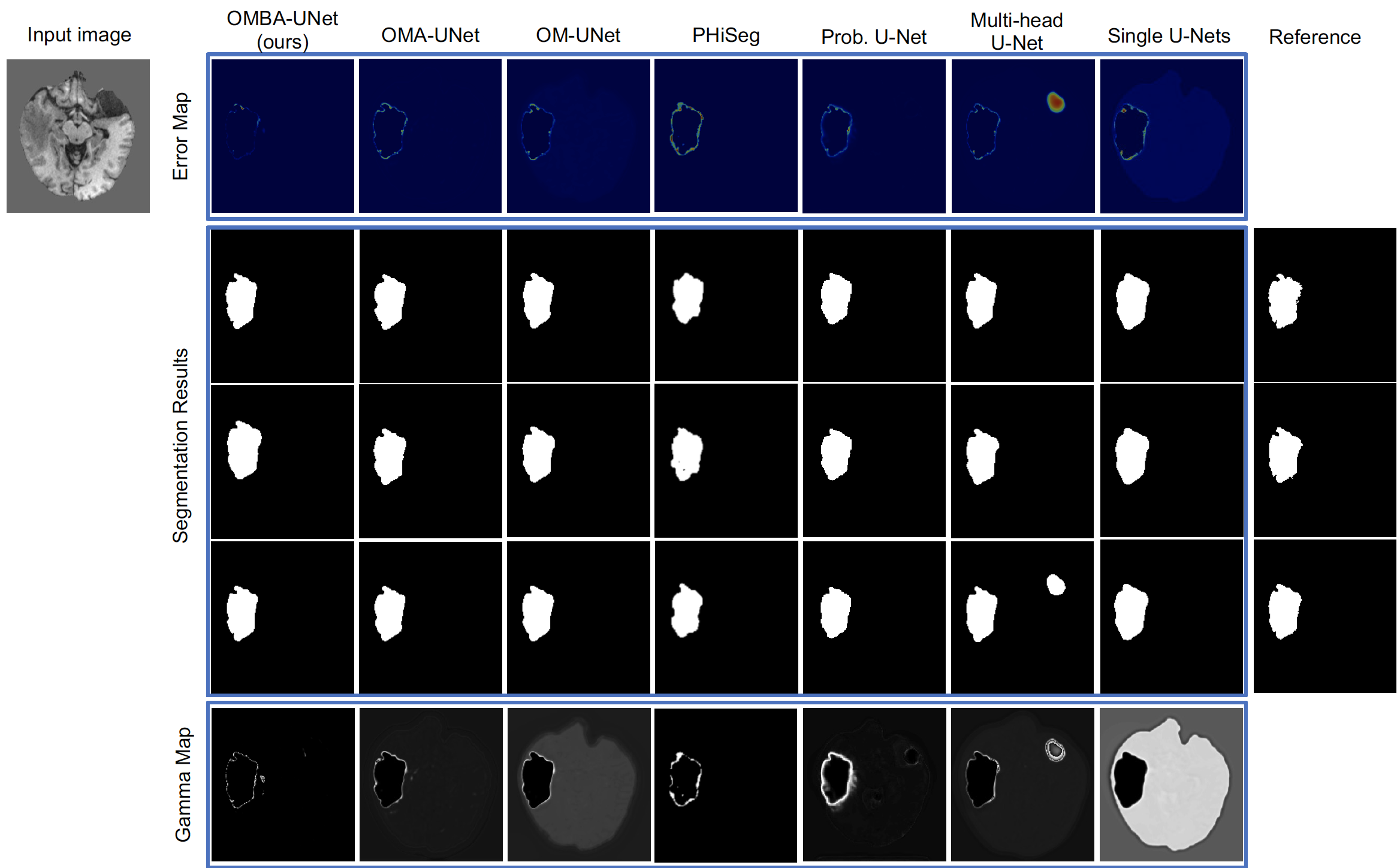}
    \caption{Illustration of the error maps $\mathbb{E}_{y, s}\left[ \text{CE}(y, s) \right]$, segmentation results, and $\gamma$-maps $\mathbb{E}\left[ \text{CE}(\bar{s}, s) \right]$ of different models trained on brain-tumor dataset with masks from three raters. The input image and corresponding annotations are shown in the top row.}
	\label{error_gamma}
\end{figure*}

Figure~\ref{attention} presents an example of input and visual intermediate attention representation. It could be seen that different regions are activated by attention coefficients. Though dynamically computed, the attention coefficients can correctly perceive the target areas, thus selecting the regions of interest in the low-level feature map and boosting the segmentation performance for different decoders. In order to demonstrate the comparison of each model, examples of a brain tumor MRI image with inter-expert disagreement are presented, in Figure~\ref{error_gamma}.  We utilize the average error maps $\mathbb{E}_{y, s}\left[ \text{CE}(y, s) \right]$ and average $\gamma-$maps, $\mathbb{E}_{\bar{s}, s}\left[ \text{CE}(\bar{s}, s) \right]$, where $\text{CE}$ is the cross-entropy, similar to \citep{baumgartner2019phiseg,gantenbein2020revphiseg} to better visualize the results in Figure~\ref{error_gamma}. Table~\ref{tab:results_Dice}.

\section{Ablation Study}


\subsection{Comparison with Bayesian Ensemble Networks}
To highlight the difference between our proposed architecture, a Bayesian neural network, and an ensemble of neural networks, in this section, we compare \emph{OMBA-UNet} with a number of approaches under both $D^2_{GED}$ and Q-score. In Table~\ref{tab:ensemble_compar}, the comparison between the four architectures is provided: \\
1) \textit{U-Net Ensemble}: Each U-Net corresponds to one decoder and each U-Net is trained independently; the final output is the average among all the U-Nets. \\
2) \textit{Bayesian U-Net Ensemble}: A U-Net ensemble method with Bayesian inference in each layer; the final output is the average among all the U-Nets. \\
3) \textit{Bayesian U-Net-AG Ensemble}: A Bayesian U-Net Ensemble method with attention gate between skip-connection,  \\
4) \textit{OMBA-UNet}:  An OM-UNet plus attention gate between skip-connection and Bayesian inference is added in every layer to sample the weight. \\
\newline
In Table~\ref{tab:ensemble_compar}, by comparing \emph{U-Net Ensemble} and \emph{Bayesian U-Net Ensemble}, the average Q-score is dramatically improved, although average $D^2_{GED}$ is slightly decreased. It indicates that Bayesian modeling does capture the inter-rater distribution more efficiently comparing the ones without, similar to what we discussed above. By comparing \emph{Bayesian U-Net Ensemble} and \emph{Bayesian U-Net-AG Ensemble} in Table~\ref{tab:ensemble_compar}, we notice that average $D^2_{GED}$ improves. This indicates the attention mechanism by enhancing the intermediate representation from each subset makes the predictive results more diverse even though with Bayesian modeling. 

More importantly, we find that \emph{OMBA-UNet} slightly outperforms \emph{Bayesian U-Net-AG Ensemble} on both average Q-score and $D^2_{GED}$. On 4 out of 7 tasks, the difference between these two models is quite small, but on two tasks, \emph{Bayesian U-Net-AG Ensemble} model even outperforms \emph{OMBA-UNet} (Kidney task1 and Prostate task1) by a large margin under $D^2_{GED}$. However, \emph{Bayesian U-Net-AG Ensemble}'s parameter number is much larger than \emph{OMBA-UNet} in Table~\ref{tab:model_parameter}. From a balanced perspective, \emph{OMBA-UNet} with fewer parameters, achieves similar or sometimes better performance than \emph{Bayesian U-Net-AG Ensemble}. 

\begin{table}[]
\centering
\caption{The parameters of all models in brain-growth segmentation task. Under LIDC-IDRI test dataset, the inference time of a single data sample. The number is rounded to four decimal places. OMBA-UNet samples 50 times for a single data sample.}
\begin{tabular}{l|c |c }
\toprule
Models                     & Parameters & Inference Time (second) \\ \midrule
Bayesian U-Net-AG Ensemble & 113M   & -    \\ \hline
Bayesian U-Net Ensemble    & 110M   & -     \\ \hline
\textbf{OMBA-UNet (ours)}   & \textbf{73M} & \textbf{0.3422} \\\hline
U-Net Ensemble             & 69M  & 0.0290      \\ \hline
OM-UNet                 & 42M   & 0.0245    \\ \hline
OMA-UNet              & 41M    & 0.0326   \\ \hline
PHiSeg             & 19M     & 0.0533 \\ \hline
Probabilistic U-Net         & 18M  & 0.0431 \\ \hline
Vanilla U-Net              & 17M & 0.0211 \\ \bottomrule
\end{tabular}
\label{tab:model_parameter}
\end{table}

\section{Discussion and Conclusion}

{
In this paper, we focus on a very challenging problem -- modeling inter-rater uncertainty. To solve this problem, we proposed a novel framework. We first presented a \emph{one-encoder-m-decoder} architecture choice to individually learn the annotation style of multiple raters while share the image feature in one encoder. We further introduce rater-specific attention module to enhance the intermediate rater-specific representation in each decoder. Moreover, we introduce Bayesian layers in our architecture to prevent the model from over-fitting on small datasets and capture the inter-rater distribution efficiently. 
When interpreting the results, we demonstrate that \emph{four} evaluation metrics consider different aspects of inter-rater uncertainty modeling. We verify proposed \emph{OM-UNet} on a newly \emph{open-sourced} rater-aligned liver dataset as well as on a synthetic LIDC-IDRI dataset. Manifested by the state-of-the-art performance on diverse segmentation tasks, our work demonstrates a promising avenue to capture inter-rater uncertainty by combining architecture choice and rater-specific attention.

One limitation is that the proposed method is relatively computationally expensive compared to U-Net Ensemble (74M vs. 69M) and significant longer inference time (0.3422 s vs. 0.0290 s) from Table~\ref{tab:model_parameter}. However, as argued in recent work \citep{le2021mc}, there is no free lunch for the variational inference to capture uncertainty. We believe that the same happens for our architecture and inference choices that could accurately model the inter-rater uncertainty.

More interestingly, we find that only considering Dice score and $D^2_{GED}$ like \citep{kohl2018probabilistic, baumgartner2019phiseg} is insufficient to reveal the actual inter-rater distribution. Therefore, in our work, we further introduce Diversity and Similarity metrics from $D^2_{GED}$ to inflect the inter-rater distribution. Even with two metrics, it is still hard to tell whether or not the model \emph{approximate} the underlying inter-rater distribution. Developing a more relabel metric to achieve this is needed. For example, a latent space-based divergence could be explored to compare the predictive distribution with the actual inter-rater distribution.

\section*{Acknowledgement}
{
This work is supported in part by National Key Research and Development Program of China (2021YFF1200800) and National Natural Science Foundation of China (Grant No. 62276121), in part by the European Research Council (ERC) under the European Union’s Horizon 2020 research and innovation programme (101045128 — iBack-epic — ERC-2021-COG).}
Bjoern Menze is supported the Helmut-Horten-Foundation. Hongwei Bran Li is supported by Forschungskredit (Grant NO.~FK-21-125) from the University of Zurich. 

\bibliographystyle{plainnat}
\bibliography{ref.bib}

\end{document}